# Modelling a DR shaft operated with pure hydrogen using a physical-chemical and CFD approach

A. Ranzani da Costa, D. Wagner, F. Patisson et D. Ablitzer

Institut Jean Lamour, INPL, CNRS, Nancy-Université, Nancy, France
fabrice.patisson@mines.inpl-nancy.fr

**Résumé** – Concevoir une nouvelle filière d'élaboration de l'acier fondée sur l'hydrogène semble prometteur si l'on considère les faibles émissions de $CO_2$ qui en résulteraient. Dans le programme ULCOS, cette filière est vue comme une option à long terme, essentiellement conditionnée par l'émergence d'une économie de l'hydrogène. Pour anticiper son possible développement, on a décidé d'évaluer le procédé de réduction directe en four à cuve par $H_2$ pur et de déterminer ses conditions opératoires optimales grâce à une approche de modélisation mathématique, complétée par des expérimentations adaptées. Nous avons ainsi spécifiquement développé un modèle mathématique, appelé REDUCTOR, qui simule le procédé et prédit ses performances. Ce modèle est fondé sur une description mathématique détaillée des phénomènes physico-chimiques et thermiques réels. En particulier, la cinétique de réduction est tirée d'essais expérimentaux en laboratoire. La version actuelle du modèle ne tient compte que de la réaction par l'hydrogène mais une extension à la réduction par CO est prévue de manière à ce que ce modèle puisse aussi être utilisé pour simuler et optimiser les procédés de réduction directe existant actuellement. Les premiers résultats ont confirmé que la réduction par $H_2$ était plus rapide que par CO, ce qui permet d'envisager un réacteur sous $H_2$ bien plus compact que les fours actuels des procédés MIDREX ou HYL.

**Abstract** – The hydrogen-based route could be a valuable way to produce steel considering its low carbon dioxide emissions. In ULCOS, it is regarded as a long-term option, largely dependent on the emergence of a hydrogen economy. To anticipate its possible development, it was decided to check the feasibility of using 100% $H_2$ in a Direct Reduction shaft furnace and to determine the best operating conditions, through appropriate experimental and modelling work. We developed from scratch a new model, called REDUCTOR, for simulating this process and predicting its performance. This sophisticated numerical model is based on the mathematical description of the detailed physical, chemical and thermal phenomena occurring. In particular, kinetics were derived from experiments. The current version is suited to the reduction with pure hydrogen, but an extension of the model to CO is planned so that it will also be adapted to the simulation and optimisation of the current DR processes. First results have confirmed that the reduction with hydrogen is much faster than that with CO, making it possible to design a hydrogen-operated shaft reactor quite smaller than current MIDREX and HYL.

## Introduction

Des procédés sidérurgiques tout à fait innovants sont analysés dans le cadre du programme européen ULCOS (Ultra low $CO_2$ steelmaking) [1], avec l'objectif de réduire d'au moins 50% les émissions de $CO_2$ par rapport au niveau actuel d'émission des usines intégrées, qui est de 1844 kg of $CO_2$/t de bobines laminées à chaud. L'utilisation d'hydrogène comme agent réducteur au lieu de CO, avec une production associée de vapeur d'eau au lieu de $CO_2$, a été étudiée dans ULCOS par le groupe SP4 (Hydrogen-based steelmaking) puis par le groupe SP12 (Advanced direct reduction), ainsi que par d'autres équipes dans le monde [2]. La filière de fabrication d'acier fondée sur l'hydrogène recommandée par le SP4 est schématisée par la figure 1. Ses performances en termes d'émission de $CO_2$ sont prometteuses: moins de 300 kg $CO_2$/t de bobines, dans le cas où l'hydrogène est produit par électrolyse de l'eau avec de l'électricité d'origine nucléaire ou hydraulique, et ce sans intégrer de capture du $CO_2$ [3].

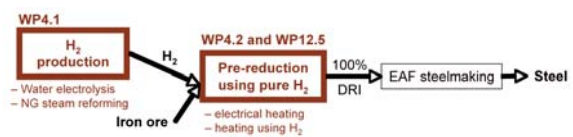

**Figure 1.** Filière hydrogène pour la fabrication d'acier /*Hydrogen-based route to steel*

Bien sûr, le développement éventuel d'une telle filière dans l'avenir est très lié à l'émergence de ce qu'il est convenu d'appeler l'économie de l'hydrogène, dans l'hypothèse d'une forte demande de ce gaz par d'autres secteurs industriels comme l'énergie ou les transports. On se place ici dans cette hypothèse, où $H_2$ est produit en grandes quantités, à un coût compétitif, et en émettant peu de $CO_2$. Nous pensons que les sidérurgistes doivent anticiper une telle éventualité et être prêts à faire dans ce cas un usage massif de l'hydrogène disponible. C'est ce qui nous a motivés à entreprendre le présent travail de recherche, consacré au procédé de réduction directe en four à cuve par $H_2$ pur et à la détermination de ses conditions opératoires optimales, à partir d'une approche combinant modélisation mathématique et expérimentation.

## Expériences

Bien que la réaction de réduction du minerai de fer par l'hydrogène ait été beaucoup étudiée dans les années 1970-80 (cf. p. ex. [4-7]), peu de publications fournissent des lois cinétiques relatives à la réduction de boulettes industrielles qui soient directement utilisables pour une modélisation mathématique. Pour préciser les mécanismes réactionnels et obtenir des données cinétiques pour la modélisation du procédé, nous avons entrepris une campagne d'expériences de thermogravimétrie, complétée par des caractérisations des échantillons par diffraction des rayons X, microscopie électronique à balayage (MEB) et spectrométrie Mössbauer

On a utilisé une thermobalance modèle TAG24 (Setaram) à deux fours symétriques, dispositif qui améliore la précision de la pesée en éliminant par construction les effets parasites des forces de traînée et de la poussée d'Archimède. Un générateur d'humidité a été installé pour pouvoir ajouter de la vapeur d'eau à l'hydrogène. Les échantillons étudiés étaient des petits cubes d'hématite (5 mm de côté) taillés dans des boulettes CVRD industrielles. Il n'était pas possible d'étudier les boulettes entières dans cette balance du fait d'une limitation sur la perte de masse totale mesurable. Les expériences de réduction ont été faites en conditions isothermes. A chaque essai, un cube unique était enroulé d'un fil de platine et suspendu au fléau de la balance. Quand la température désirée était atteinte, on ouvrait la vanne d'hydrogène et le courant d'hydrogène était maintenu jusqu'à la fin de la réduction. Les compositions gazeuses employées ont été 100 % $H_2$ et 60 % (vol.) $H_2$ dans He. Dans certains cas, on a aussi ajouté quelques (jusque 4%) pourcents d'eau dans le mélange gazeux. Les températures ont varié entre 600 et 990 °C. L'effet de la température (figure 2) est complexe et mérite d'être discuté.

Aux températures inférieures à 800 °C, plus la température augmente, plus la réaction est rapide, sauf à 700 °C. Au-dessus de 800 °C, une température plus élevée commence par provoquer une réaction plus rapide mais, en fin de conversion, la réaction ralentit et se poursuit plus lentement, conduisant à un temps de conversion complète globalement plus long qu'à 800 °C. Ainsi, au vu de ces essais, 800 °C apparaît comme la température optimum pour une conversion totale en un temps minimum.

Une série d'expériences interrompues, lors desquelles l'arrivée d'hydrogène était coupée avant la fin de la réduction, a été réalisée en vue de caractériser les solides à différents degrés de conversion. Les échantillons partiellement réduits furent observés au MEB et analysés par spectrométrie Mössbauer. Ces expériences ont révélé les détails du déroulement de la réaction, au travers de la formation des différents oxydes intermédiaires et de l'évolution morphologique des solides (figure 3).

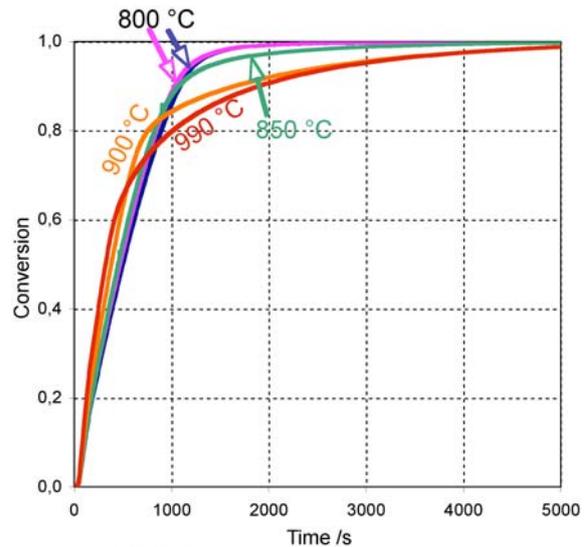

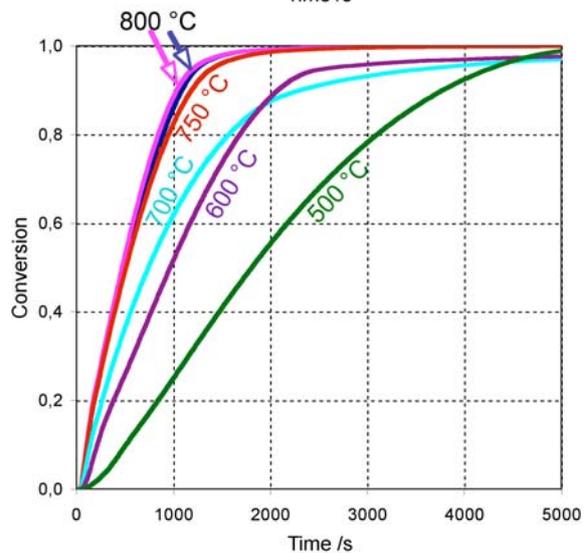

**Figure 2.** Influence de la température sur la cinétique de réduction de cubes d'hématite, avec 60 % $H_2$ dans He / *Influence of temperature on the reduction curves of hematite cubes, under 60 % $H_2$ in He*

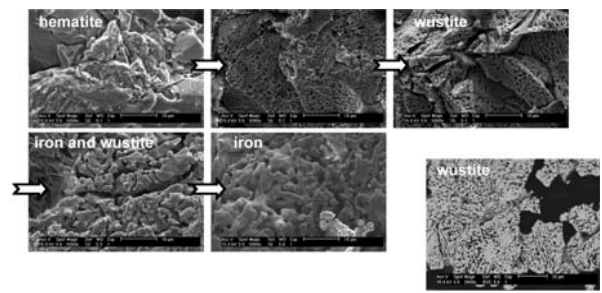

**Figure 3.** Evolution morphologique du solide lors de la réduction de cubes d'hématite à 800 °C, 60 % $H_2$ dans He / *Morphological evolution of the solid during the reduction of hematite cubes at 800 °C, 60 % $H_2$ in He*

Les boulettes d'hématite sont constituées de gros grains denses, séparés des de fins espaces. La structure granulaire change assez peu de l'hématite à la magnétite et à la wüstite. On note surtout l'apparition et la croissance de petits pores à la surface des grains. Au stade de la wüstite, cependant, l'observation de sections des grains (photo en bas à droite) montre que ceux-ci sont devenus très poreux et comme subdivisés en plus petits grains, que nous nommerons cristallites. De la wüstite au fer, la structure change énormément. Le fer forme des grains plus gros, plus lisses et plus denses, mais la porosité inter-granulaire demeure élevée.

### Modèle d'une boulette unique

Un modèle mathématique qui simule la réduction d'une seule boulette a été conçu pour faire le lien entre les expériences et le modèle de réacteur multi-particulaire. Utilisé indépendamment, il aide à l'interprétation et à la simulation des essais expérimentaux. Utilisé comme sous-modèle du modèle de réacteur, il prédit la vitesse de réaction en fonction des conditions locales de réduction (température et composition du gaz). Ce modèle d'une boulette unique a été conçu en tenant compte des observations expérimentales. La boulette (diamètre typique 12 mm) est supposée constituée de grains (diamètre 25 μm), avec une porosité inter-granulaire d'environ 10 %. Les grains sont eux-mêmes poreux (porosité intra-granulaire de 50 %) et, dans le cas de la wüstite, composés de cristallites (2 μm) denses séparées par des pores très fins.

Pour que la description mathématique reste simple, ce qui était nécessaire en vue de l'intégration de ce modèle dans REDUCTOR, nous avons retenu le concept dit de la loi des temps caractéristiques additifs [8] pour calculer la vitesse globale de la réaction. Cette loi considère que les différentes étapes du transport de matière interviennent en série, ce qui fait que leurs résistances, représentées par des temps caractéristiques, s'ajoutent. Une relation unique, compliquée mais analytique, exprime alors la vitesse instantanée de la réaction globale. Cette loi n'est qu'approchée mais sa validité a été démontrée pour plusieurs systèmes gaz-solide [9]. Son principal intérêt est de pouvoir rendre compte des régimes cinétiques mixtes au travers d'une seule expression, dont le calcul dans un sous-programme d'un code de réacteur reste rapide.

Les expressions des différents temps caractéristiques et des vitesses correspondantes sont données dans [10]. Les phénomènes suivants sont pris en compte : transfert externe de $H_2$ et $H_2O$ à travers la couche limite entourant la boulette ; diffusion gazeuse dans les pores inter-granulaires, intra-granulaires et dans ceux de la couche de fer autour des cristallites lorsque celle-ci est poreuse, ces deux dernières processus incluant la diffusion de Knudsen ; la diffusion en phase solide de l'oxygène à travers la couche de fer lorsque celle-ci est dense ; les trois réactions hétérogènes de réduction

$$3Fe_2O_3 + H_2 = 2Fe_3O_4 + H_2O$$
$$Fe_3O_4 + \frac{16}{19}H_2 = \frac{60}{19}Fe_{0,95}O + \frac{16}{19}H_2O$$
$$Fe_{0,95}O + H_2 = 0,95\,Fe + H_2O$$

et enfin le possible frittage de la phase fer.

La description de ce dernier phénomène est une composante originale de notre modèle, introduite pour rendre compte de deux observations expérimentales : premièrement, le ralentissement cinétique remarqué (figure 2, haut) en fin de conversion et aux températures supérieures à 850 °C et, deuxièmement, le grossissement et la densification des grains de fer au-dessus de cette même température. Nous expliquons ce comportement par la tendance que présente la couche de fer fraîchement formé, au niveau des cristallites, à fritter, c'est-à-dire à réduire sa surface totale. Il s'ensuit une diminution des pores intra-cristallites, ce qui ralentit la diffusion gazeuse à travers ces pores, pouvant aller jusqu'à leur disparition, ce qui impose que la réduction ne peut ensuite se poursuivre qu'en faisant appel à de la diffusion en phase solide, plus lente. Les équations correspondantes sont également données dans [10].

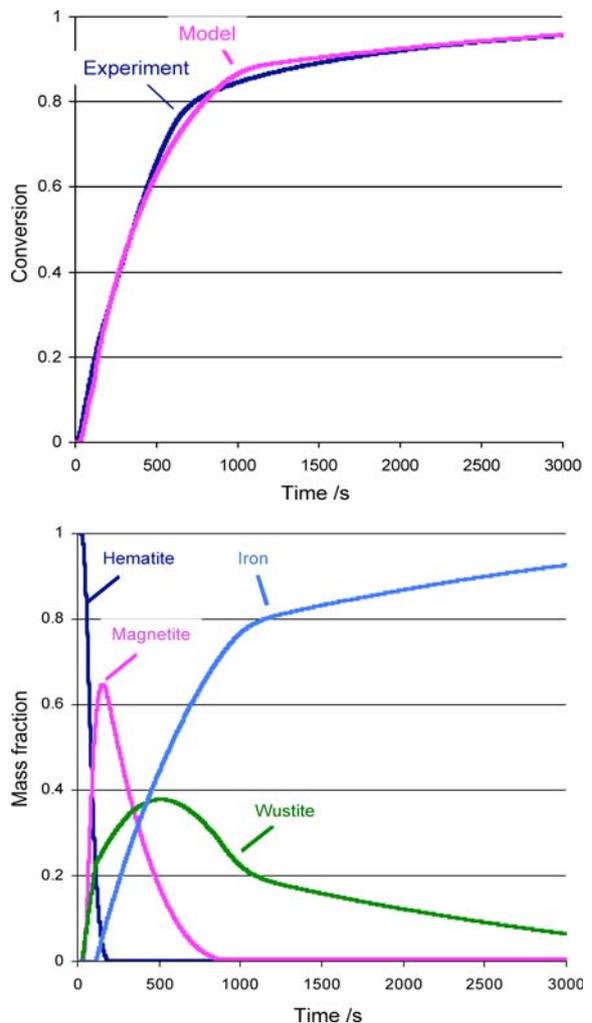

**Figure 4.** Résultats du modèle d'une boulette unique, 900°C, 60 % $H_2$ dans He. Haut : comparaison à l'expérience ; bas : évolution de la fraction massique des différents oxydes et du fer
/Results of the single pellet model, 900°C, 60 % $H_2$ in He. Top: comparison with experiment; bottom: evolution of the mass fractions of the different oxides and iron

Les principaux résultats du modèle de boulette unique sont donnés figure 4. La comparaison avec l'expérience (graphique du haut) montre que le modèle simule très bien l'expérience. On note sur les deux courbes le ralentissement à 80 % de conversion. La succession des trois réactions (hématite → magnétite → wüstite → fer) est illustrée sur le graphique du bas. L'hématite disparaît rapidement, alors que la réduction de la wüstite est l'étape la plus longue.

### REDUCTOR : modèle d'un four à cuve de réduction directe par $H_2$ pur

Pour prédire la performance d'un procédé qui n'existe pas encore, la modélisation mathématique nous a semblé la meilleure approche. Elle donne accès aux principales variables (températures, compositions et vitesses du solide et du gaz, vitesses des réactions, degré de réduction, etc.) localement. Nous avons opté pour une modélisation de type mécanique des fluides numérique, fondée sur la résolution numérique des bilans locaux de matière, énergie et quantité de mouvement. Au-delà des bilans globaux, également fournis par les logiciels et modèles de flow-sheeting, cette approche révèle le fonctionnement interne et détaillé d'un réacteur. De plus, l'influence des conditions opératoires et des paramètres physiques peut être analysée de manière quantitative et des conditions optimales peuvent être dégagées.

REDUCTOR est un modèle numérique bi-dimensionnel (en coordonnées cylindriques) en régime permanent. Les principles équations considérées sont les bilans locaux de matière, énergie et quantité de mouvement écrits ci-dessous. Les notations sont les suivantes : $r$ = rayon, $z$ = hauteur, $c_t$ = concentration totale du gaz, $x_i$ = titre molaire de $i$ dans le gaz, $u_g$ = vitesse en fût vide du gaz, $D_a$ et $D_r$ = coefficients de dispersion axial et radial, $r_1$, $r_2$, $r_3$ = vitesses (mol s$^{-1}$ m$^{-3}$) des trois réactions de réduction : hématite → magnétite, magnétite → wüstite et wüstite → fer, $w_j$ = titre massique de $j$ dans le solide, $M_i$ = masse molaire de $i$, $\rho_g$ = masse volumique du gaz, $\rho_s$ = masse volumique apparente du lit solide, $c_{pg}$, $c_{ps}$ = chaleurs massiques du gaz et du solide, $c^*_{p,i}$ = chaleur molaire de $i$, $\lambda_g$ = conductivité thermique du gaz, $T_g$, $T_s$ = températures du gaz et du solide, $a_g$ = aire spécifique du lit, $h$ = coefficient d'échange thermique gaz-solide, $\lambda_{eff,a}$, $\lambda_{eff,r}$ = conductivités axiale et radiale du lit solide, $\Delta_r H_i$ = chaleur de la réaction $i$, $p$ = pression du gaz, $\varepsilon$ = porosité inter-particulaire du lit, $d_p$ = diamètre des boulettes, $\mu_g$ = viscosité du gaz.

Bilans matières gaz :

$$\frac{1}{r}\frac{\partial(rc_t x_i u_{gr})}{\partial r} + \frac{\partial(c_t x_i u_{gz})}{\partial z} = \frac{1}{r}\frac{\partial}{\partial r}\left(rc_t D_r \frac{\partial x_i}{\partial r}\right) + \frac{\partial}{\partial z}\left(c_t D_a \frac{\partial x_i}{\partial z}\right) + S_i$$

avec $i = H_2$ ou $H_2O$ et $-S_{H_2} = S_{H_2O} = r_1 + \frac{16}{19}r_2 + r_3$

Bilans matières solides :

$$-\frac{\partial(\rho_s u_s w_j)}{\partial z} = S_j$$

avec $j = Fe_2O_3$, $Fe_3O_4$, $Fe_{0,95}O$, ou $Fe$, et

$S_{Fe_2O_3} = -3 M_{Fe_2O_3} r_1$

$S_{Fe_3O_4} = M_{Fe_3O_4}(2r_1 - r_2)$

$S_{Fe_{0,95}O} = M_{Fe_{0,95}O}\left(\frac{60}{19}r_2 - r_3\right)$

$S_{Fe} = 0,95 M_{Fe} r_3$

Bilan thermique du gaz :

$$\rho_g c_{pg}\left(u_{gr}\frac{\partial T_g}{\partial r} + u_{gz}\frac{\partial T_g}{\partial z}\right) = \frac{1}{r}\frac{\partial}{\partial r}\left(r\lambda_g \frac{\partial T_g}{\partial r}\right) + \frac{\partial}{\partial z}\left(\lambda_g \frac{\partial T_g}{\partial z}\right)$$
$$+ a_g h(T_s - T_g) + \left(r_1 + \frac{16}{19}r_2 + r_3\right)\int_{T_g}^{T_s}\left(c^*_{P,H_2O} - c^*_{P,H_2}\right)dT$$

Bilan thermique du solide :

$$-\rho_s u_s c_{ps}\frac{\partial T_s}{\partial z} = \frac{1}{r}\frac{\partial}{\partial r}\left(r\lambda_{eff,r}\frac{\partial T_s}{\partial r}\right) + \frac{\partial}{\partial z}\left(\lambda_{eff,a}\frac{\partial T_s}{\partial z}\right) +$$
$$a_g h(T_g - T_s) + r_1(-\Delta_r H_1) + r_2(-\Delta_r H_2) + r_3(-\Delta_r H_3)$$

Bilan de quantité de mouvement combiné à l'équation de la continuité :

$$\frac{1}{r}\frac{\partial}{\partial r}\left(r\frac{c_t}{K}\frac{\partial p}{\partial r}\right) + \frac{\partial}{\partial z}\left(\frac{c_t}{K}\frac{\partial p}{\partial z}\right) = 0$$

avec $K = 150\frac{(1-\varepsilon)^2}{\varepsilon^3 d_p^2}\mu_g + 1,75\frac{(1-\varepsilon)}{\varepsilon^3 d_p}\rho_g u_g$

Les vitesses des trois réactions sont fournies par le modèle de la boulette unique, avec les paramètres cinétiques tirés des expériences.

A ces équations aux dérivées partielles s'ajoutent les conditions aux limites (température et composition connues aux entrées, flux nuls sur l'axe de symétrie et à la paroi, etc.). Pour plus de détails, ainsi que pour les corrélations utilisées pour calculer les paramètres physico-chimiques et thermiques, cf. [10].

La résolution numérique s'appuie sur la méthode des volumes finis [11] : les équations aux dérivées partielles sont discrétisées, et le système d'équations algébriques qui en résulte est résolu par itérations successives en suivant l'algorithme de Gauss-Seidel. Le maillage comporte 201 (verticalement) × 21 (radialement) cellules. Le code est écrit en Fortran 90 et, exécuté sur un cluster de PC (16 processeurs AMD Opteron), dure typiquement 40 h.

Nous avons commencé par une simulation de référence. Le domaine de calcul correspond à uniquement la partie cylindrique d'un four à cuve, c'est-à-dire la zone de réduction, au-dessus de l'entrée latérale du gaz réactif. La figure 5 précise la géométrie du four modélisé et les entrées gaz et solide. Le débit de solide correspondant à une production an-

A. Ranzani da Costa, D. Wagner, F. Patisson and D. Ablitzer

nuelle de 1 Mt$_{Fe}$ an$^{-1}$, et le débit total de gaz (3734 mol s$^{-1}$) est égal à 3,8 fois le débit stœchiométrique.

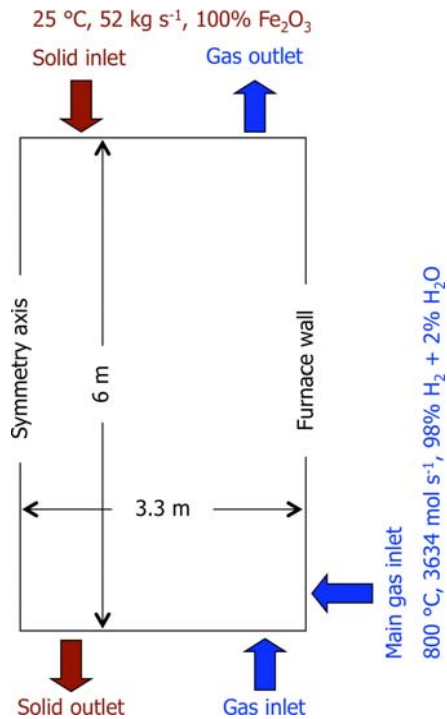

**Figure 5.** Conditions du calcul de référence / *Conditions of computation in the reference case*

La figure 6 présente les cartes des titres massiques caculés des différents oxydes et du fer. Comme avec le modèle d'une boulette unique, la réduction de l'hématite et celle de la magnétite sont rapides. Celle de la wüstite l'est moins mais, à $z=2$ m, tous les oxydes sont complètement convertis en fer quelle que soit la position radiale (pour cette simulation). C'est un résultat important puisqu'il montre qu'une hauteur d'environ 4 m est suffisante pour atteindre 100 % Fe, alors que les fours de réduction directe fonctionnant au gaz de synthèse (CO+H$_2$) ont une hauteur typique de 9 m (pour un même diamètre de 6,6 m – four MIDREX – comme dans cette simulation) et ne permettent d'atteindre que 92 % Fe.

Nous avons réalisé d'autres simulations pour étudier l'influence des conditions opératoires et des paramètres physiques du modèle. Nous montrons ici l'effet de la teneur en eau dans le gaz d'entrée et celui de la taille des boulettes (figures 7 et 8).

L'utilisation d'un gaz plus humide diminue la force motrice des réactions (réversibles) et donc leur vitesses. La carte des titres en fer obtenus quand le gaz introduit contient 10 % d'eau (figure 7, droite), à comparer à celle du cas de référence, dans lequel le gaz injecté ne contenait que 2 % d'eau, montre que la réduction n'est pas complète, en particulier dans la zone centrale.

Pour ce qui est de l'effet de la taille des boulettes, un diamètre de 24 mm conduit à une conversion incomplète (75 % de fer en moyenne en sortie), alors qu'avec des boulettes de 6 mm, une réduction totale est atteinte seulement 2 m en aval de l'entrée, résultat à comparer aux 4 m nécessaires dans le cas de référence avec des boulettes de 12 mm. Par conséquent, utiliser des boulettes plus petites semble une piste tout à fait intéressante pour la pratique industrielle.

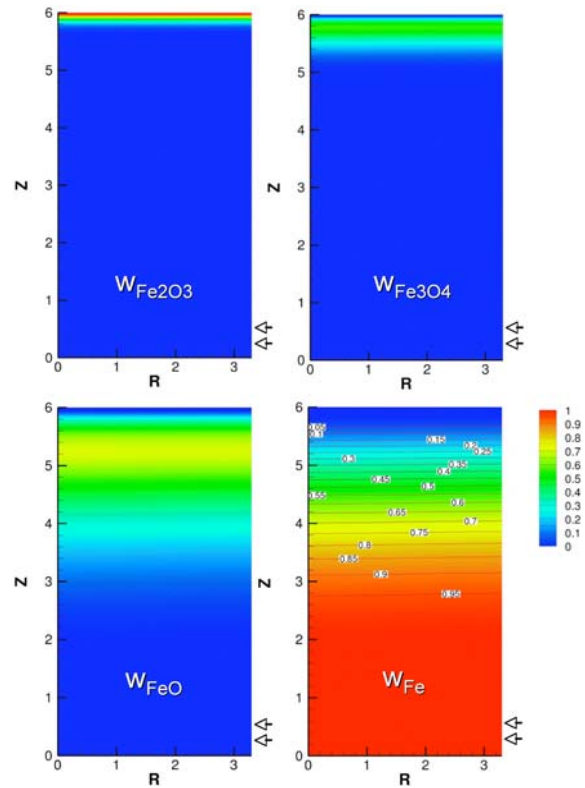

**Figure 6.** Titres massiques des solides calculés dans le cas de référence : gaz à 800 °C, 98 %H$_2$, $d_p$ =12 mm / *Calculated mass fractions of the solid species in the reference case: gas in 800 °C, 98 %H$_2$, $d_p$ =12 mm*

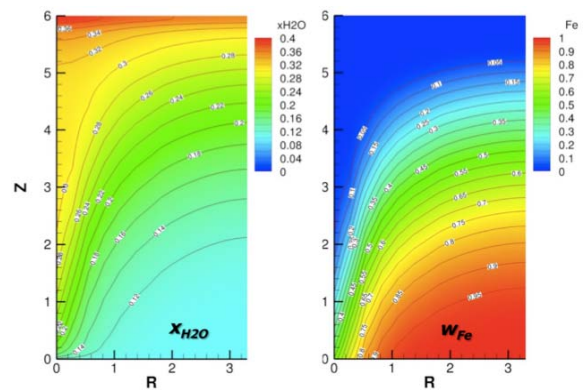

**Figure 7.** Titre molaire en vapeur d'eau dans le gaz (à gauche) et titre massique en fer dans le solide (à droite) quand le gaz réactif introduit contient 10% H$_2$O / *Calculated molar fractions of water vapour in the gas (left) and mass fraction of iron in the solid (right), when introducing 10% H$_2$O in the gas*

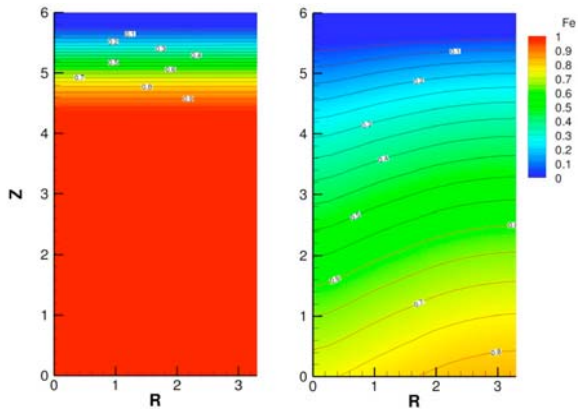

**Figure 8.** Titres massiques en fer calculés pour différentes tailles des boulettes : $d_p$=6 mm (à gauche) and $d_p$ =24 mm (à droite)
/*Calculated mass fractions of iron for different pellet sizes: $d_p$=6 mm (left) and $d_p$ =24 mm (right)*

## Conclusion

Nous avons développé un nouveau modèle (REDUCTOR), bidimensionnel en régime permanent, qui simule un four à cuve de réduction directe du minerai de fer fonctionnant sous hydrogène. Ce modèle décrit l'écoulement du gaz et du solide, les transferts de matière et de chaleur, et la réduction en trois étapes de l'hématite en fer. Les cinétiques des réactions sont calculées par un modèle spécifique de conversion d'une boulette unique, basé sur la loi des temps caractéristiques additifs et nos résultats expérimentaux. Les premières simulations confirment les atouts de la réduction directe par l'hydrogène : les émissions de $CO_2$ du réacteur lui-même sont quasi nulles et, du fait de la meilleure réactivité de $H_2$ (comparé à CO), une réduction complète en fer métallique peut être obtenue dans un réacteur bien plus compact (typiquement deux fois plus petit) que les fours de réduction directe classiques MIDREX ou HYL. Notre prochain travail sur ce modèle consistera à prendre en compte la réduction par CO afin de pouvoir simuler les procédés existants et de valider le modèle par comparaison à des données industrielles.

A. Ranzani da Costa, D. Wagner, F. Patisson and D. Ablitzer